\documentclass[useAMS,usenatbib]{mn2e}
\usepackage{graphicx} 
\usepackage{lscape}
\usepackage{wasysym} 
\newcommand{\reff}[1]{{ #1}}
\title[Warm \emph{Spitzer} Occultation Photometry of WASP-26b ]{Warm \emph{Spitzer} Occultation Photometry of WASP-26b at 3.6$\micron$ and 4.5$\micron$}

\author[D.P.Mahtani et al.]{D.P. Mahtani$^{1}$\thanks{E-mail:
d.p.mahtani@keele.ac.uk}, 
P.F.L. Maxted$^{1}$,  
D.R. Anderson$^{1}$, 
A.M.S. Smith$^{1,2}$, 
B. Smalley$^{1}$ 
\newauthor J. Tregloan-Reed$^{1}$, 
J.  Southworth$^{1}$, 
N. Madhusudhan$^{3}$,
A. Collier Cameron$^{4}$, 
\newauthor M. Gillon$^{5}$,
J. Harrington$^{6}$,
C. Hellier$^{1}$,
D. Pollacco$^{7}$,
D. Queloz$^{8}$,
\newauthor A.H.M.J. Triaud$^{8}$,
R.G. West$^{9}$\\
$^{1}$ Astrophysics Group, Keele University Staffordshire,ST5 5BG \\ 
$^{2}$ Current address: N. Copernicus Astronomical
Centre, Polish Academy of Sciences, Bartycka 18, 00-716 Warsaw,
Poland\\
$^{3}$ Department of Physics and Department of Astronomy, Yale University, New Haven, CT, 06520-8101, USA\\
$^{4}$ SUPA, School of Physics and Astronomy, University of St. Andrews, North Haugh, Fife KY16 9SS\\
$^{5}$ Institut d'Astrophysique et de G\'eophysique, Universit\'e de
Li\`ege, All\'ee du 6 Ao\^ut, 17, Bat. B5C, Li\`ege 1, Belgium\\
$^{6}$ Planetary Sciences Group, Department of Physics, University of Central Florida, Orlando, FL 32816-2385, USA\\
$^{7}$ Department of Physics, University of Warwick, Gibbet Hill Road, Coventry CV4 7AL, UK\\
$^{8}$ Observatoire astronomique de l'Universit\'e de Gen\`eve 51 ch. des Maillettes, 1290 Sauverny, Switzerland \\
$^{9}$ Department of Physics and Astronomy, University of Leicester, Leicester, LE1 7RH, UK\\}

\begin{document}

\maketitle

\label{firstpage}

\begin{abstract}

We present new warm \emph{Spitzer} occultation photometry of WASP-26 at
$3.6\micron$ and $4.5\micron$ along with new transit photometry taken in the
\emph{g,r} and \emph{i} bands. We report the first detection of the
occultation of WASP-26b, with occultation depths at $3.6\mu$m and $4.5\mu$m of
$0.00126 \pm 0.00013$ and $0.00149 \pm 0.00016$ corresponding to brightness
temperatures of $1825\pm80$K and $1725\pm89$K, respectively. We find that the
eccentricity of the orbit is consistent with a circular orbit at the $1\sigma$
level ($e=0.0028 ^{+ 0.0097}_{- 0.0022}$, $3\sigma$ upper limit $e<0.04$).
According to the activity-inversion relation of \citet{Knutson2010}, WASP-26b
is predicted to host a thermal inversion. The brightness temperatures deduced
from the eclipse depths are consistent with an isothermal atmosphere, although
it is within the uncertainties that the planet may host a weak thermal
inversion. The data are equally well fit by atmospheric models with or without
a thermal inversion. We find that variation in activity of solar-like stars
does not change enough over the time-scales of months or years to change the
interpretation of the \citet{Knutson2010} activity-inversion relation,
provided that the measured activity level is averaged over several nights.
Further data are required to fully constrain the thermal structure of the
atmosphere because the planet lies very close to the boundary between
atmospheres with and without a thermal inversion.


\end{abstract}


\section{Introduction}


The first detection of thermal emission from an exoplanet was reported by
\citet{deming2005} and \cite{char2005}. The teams observed the secondary
eclipse of HD209458 and TrES-1 using the \emph{Spitzer} Space Telescope.
Secondary eclipses of many other exoplanets have now been observed
\citep[e.g.][] {mach2008,anderson2011b,Todorov2012}. Through the
spectrophotometry of this event, observed using \emph{Spitzer} and ground
based telescopes, we can build up the spectral energy distribution (SED) of
the irradiated hemisphere (day side) of the planet. From the SED we can
investigate the atmospheric properties of the day side of the planet.
Secondary eclipse observations made with \emph{Spitzer} have shown that some
of these exoplanets have temperature inversions
\citep{Fortney2008,Knutson2009a,Madhusudhan2010}. Thermal inversions are
thought to form when gases exist in the upper atmosphere of these exoplanets
that are efficient absorbers of optical and ultraviolet light
\citep{Fortney2008}. This absorption of radiation causes the temperature of
this region of the atmosphere to increase. Gases that have been hypothesised
to cause thermal inversions to form are titanium oxide and vanadium oxide
\citep{Spiegel2009} and sulphur compounds \citep{Zahnle2009}.
  

WASP-26b, discovered by \citet{smalley2010} with SuperWASP
\citep{pollacco2006}, is a 1 Jupiter mass  ($1M_{Jup}$) planet in a $2.8$  day orbit around a G0 type star. WASP-26 also has a common proper motion companion 15" away \citep{smalley2010}. \citet{anderson2011} conducted an investigation using the Rossiter-McLaughlin effect to determine the sky-projected spin-orbit angle of the system. However, their results were inconclusive. \citet{al2012} constrained the spin-orbit angle of the system to $\lambda = -34 ^{+36}_{-26}$$^{\circ}$. In this paper we present new warm \emph{Spitzer} and ground based photometry of WASP-26.

\section{observations}
We present \emph{Spitzer} \citep{werner2004} InfraRed Array Camera (\emph{IRAC}) \citep{Fazio2004} channel 1
(3.6$\micron$) and channel 2 (4.5$\micron$) secondary eclipse (occultation)
data taken on 2010 August 3 and 2010 September 7-8, respectively (PI: J H,
Program ID 60003). The \emph{Spitzer} data were acquired in full array mode
(256 $\times$ 256). Also presented are new full transit data taken in the
\emph{g, r} and \emph{i} bands (taken simultaneously) using the 2.2m telescope
at the Calar Alto Astronomical Observatory with the Bonn University
Simultaneous CAmera (BUSCA) on 2010 August 20. BUSCA is a 4 channel CCD
photometer with 4096 $\times$ 4096 pixels per CCD  with a plate scale of
$0.17$ arc seconds per pixel. The BUSCA transit data were obtained using
defocused photometry \citep{southworth2009,southworth2012} and BUSCA was used
with a 256 $\times$ 1400 pixel window and 2$\times$2 binning to reduce the
read out time.  Table \ref{obs} is a summary of the data that we have  used in
our analysis.

\begin{table*}
\begin{center}
\caption{Summary of data used in this analysis.}
\begin{tabular}{l|l|l|}
\hline
Observation & Dates & Publication\\
\hline
SuperWASP Lightcurves (400 - 700 nm filter) &  2008 June 30 - 2008 November 17 & \citet{smalley2010}\\  & 2009 June 28 - 2009 November 17\\ 
16 RV spectra from CORALIE & 2009 June 19 - 2009 August 22 & \citet{smalley2010}\\(1.2m Swiss Telescope,La Silla, Chile)\\
30 RV spectra from HARPS  & 2010 September 12 & \citet{anderson2011}\\(HARPS Spectrograph, ESO 3.6m telescope, La Silla, Chile) & &\\
Full transit (Pan-STARRS-z filter)  & 2009 November 18 & \citet{smalley2010}\\(2.0m Faulkes Telescope South, Siding Spring, Australia)\\
Occultation (3.6$\mu$m) & 2010 August 3 & This Paper\\ (\emph{Spitzer} channel 1)\\
Occultation (4.5$\mu$m) & 2010 August 7 - 2010 August 8 & This Paper\\ (\emph{Spitzer} channel 2)\\
Full transit (\emph{g, r} and \emph{i} band)  & 2010 August 20 & This paper\\(Calar Alto Astronomical Observatory with BUSCA, Almer\'ia, Spain)\\
\hline
\end{tabular}
\label{obs}
\end{center}
\end{table*}

\section{Data Reduction}

\subsection{Transit Data Reduction}
\begin{figure}
\includegraphics[scale=0.36]{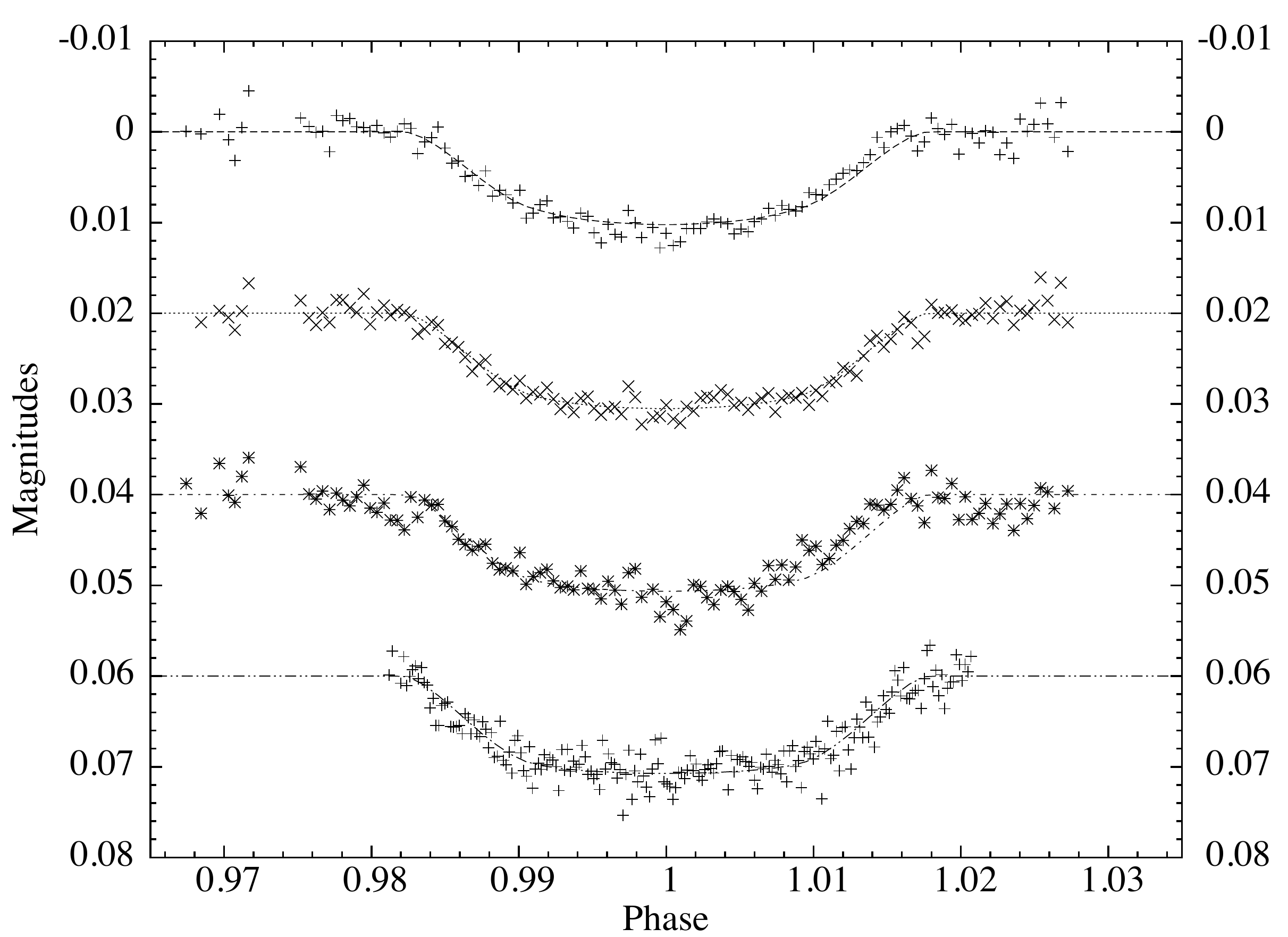}
\caption{BUSCA and FTS data with best fitting models (from the top to bottom),  \emph{g} band data, \emph{r} band data, \emph{i} band data and the FTS (z filter) data. }
\label{busca}
\end{figure}

We used an IDL implementation of DAOPHOT \citep{stetson1987} to perform
synthetic aperture photometry on our BUSCA images, as in
\citet{southworth2009}. Light curves were obtained in the \emph{g, r} and
\emph{i} bands. In all three bands one comparison star was used. 
We used a target aperture radius of 24 pixels,  a sky annulus of
inner radius 70 pixels and an outer radius 100 pixels for the \emph{g, r} and
\emph{i} bands. The wings of the PSF of the companion star do contaminate the
target aperture but the contribution to the observed flux is negligible.
Iterative outlier removal  was
used on the image values in the sky annulus to remove the effect of the
light from the wings of the companion's PSF in the sky annulus. The light
curves are shown in Figure \ref{busca}.

\subsection{\emph{Spitzer} Data Reduction}
The data reduction was conducted using the Image Reduction and Analysis Facility (IRAF)\footnote{IRAF is distributed by the National Optical Astronomy Observatories, which are operated by the Association of Universities for Research in Astronomy, Inc., under cooperative agreement with the National Science Foundation.} using the same method as \citet{anderson2011b}, described briefly below.

We convert from MJy/sr to electrons using equation (\ref{factor}), where the gain, exposure time and flux conversion factor were taken from the image headers.

\begin{equation}
\label{factor}
\rm{Factor}=\frac{\rm{Gain \times Exposure~Time}}{\rm{Flux~Conversion~Factor}} 
\end{equation}


Aperture photometry was then conducted using the PHOT procedure in IRAF, using
21 aperture radii in the range 1.5-6 pixels and with a sky annulus of inner
radius 8 pixels and outer radius 16 pixels. It was found that the stellar
companion to WASP-26 and a bad column in channel 2 data were both inside the
sky annulus. However, an iterative 3-sigma clipping was conducted which
excludes those pixels. The error on the photometry was calculated from the
photon statistics and the read out noise of the IRAC detectors. The readout
noise values were taken from the IDL program
SNIRAC\_warm.pro,\footnote{$ssc.spitzer.caltech.edu/warmmission/propkit\\/som/snirac{_{-}}warm.pro$}
the values for channel 1 and 2 are 9.87 and 9.4 electrons, respectively. The
position of the target was measured by fitting a 1-dimensional Gaussian to the
marginal distributions of flux on $x$ and $y$ image axes. For each data set
the times of mid-exposure were converted to BJD$_{\rm TDB}$ \citep{time} and
for the occultation data
the light travel time across the system ($\sim$40s) was accounted for. The
light travel time across the system was calculated using the semi-major axis
from the output of our initial MCMC (see below for details of this run) and
this time was subtracted from all the \emph{Spitzer} times.




\section{analysis}

\subsection{MCMC}

We explored the parameter space using a Markov chain Monte Carlo (MCMC)
algorithm \citep{camron2007,pollacco2008,enoch2010}. The input parameters for
the star that were used in the MCMC analysis are $\rm{T}_{\rm{eff}}=5950 \pm
100 $ and $  [Fe/H] =-0.02 \pm 0.09$ \citep{anderson2011}.  Stellar density,
which is directly constrained by the transit light curve and the spectroscopic
orbit \citep{sm2003} and the eccentricity of the orbit, is calculated from the
proposal parameter values. This is input, together with the latest values of
$\rm{T}_{\rm{eff}}$ and [Fe/H] (which are controlled by Gaussian priors) into
the empirical mass calibration of \citet{enoch2010} to obtain an estimate of
the stellar mass, $M_{\star}$.  At each step in the MCMC procedure, each
proposal parameter is perturbed from its previous value by a small, random
amount. From the proposal parameters, model light and RV curves are generated
and $\chi^2$ is calculated from their comparison with the data. A step is
accepted if $\chi^2$ (our merit function) is lower than for the previous step,
and a step with higher $\chi^2$ is accepted with probability $\exp(-\Delta
\chi^2)$. In this way, the parameter space around the optimum solution is
thoroughly explored. The value and uncertainty for each parameter are
taken as the median and central 68.3 per cent confidence interval
of the parameter's marginalised posterior probability distribution, respectively 
\citep[][]{Ford2006}. \reff{The median closely approximates the $\chi^2$ minimum for symmetric posteriors such as ours, and is more robust to noise in the case of flat minima.} Table \ref{params} show the proposal parameters of
the MCMC. We did an initial run which included all the transit photometry,
including WASP photometry, to get a good estimate of the epoch of mid-transit.
This value along with its uncertainty were used as a Bayesian prior in
subsequent MCMC runs which used all the photometry, including \emph{Spitzer},
but excluding the WASP photometry (to reduce computing time). The  transit
model used in the analysis was the small planet approximation of
\citet{mandel2002} with 4-parameter limb darkening coefficients taken from 
\citet{claret2004}. The limb darkening coefficients were determined using an initial
interpolation in $\log g_{*}$ and {[Fe/H]} and an interpolation in $T_{\rm
eff}$ at each MCMC step. The limb darkening parameters used for the best-fit
lightcurves are given in Table \ref{ld}.  For the secondary eclipse we
approximated the star and planet as two uniform discs of constant surface
brightness. We fixed the projected spin-orbit angle to the value $\lambda=0$ in our fit since the HARPS data covering
the transit are negligibly affected by the R-M effect. The fit to the optical
lightcurves (Figure \ref{busca}) shows that there is some red noise present in
the \emph{g} and \emph{i} band lightcurves. We have accounted for the small
additional uncertainty due to this noise in our quoted parameter standard
errors rather than trying to find an arbitrary model that would improve the
fit.

\begin{figure*}
	\begin{center}
		\includegraphics{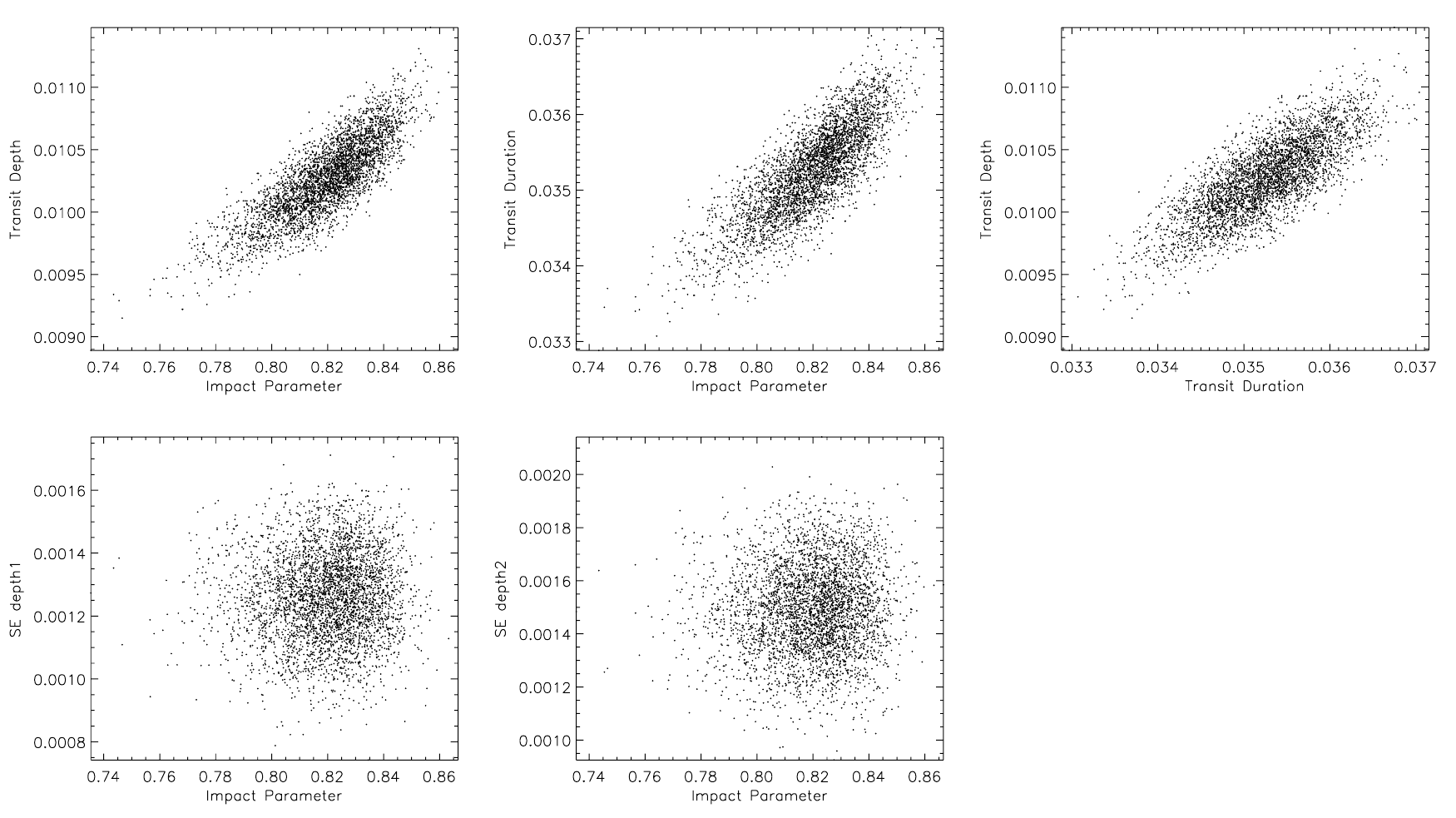}
	\end{center}
	\caption{\reff{Correlation plots for selected proposal parameters from our MCMC analysis. For clarity we have only plotted a random 2\% of the chain values.}}
	\label{fig:transcor}
\end{figure*}

\reff{We checked for any correlations in our proposal parameters and only found the correlation between transit depth, width and impact parameter often seen in ground-based lightcurves. These correlations are caused by the blurring of the second and third contact points due to limb darkening in the optical lightcurves. These correlations do not affect our secondary eclipse depth measurements. These correlations are shown in Figure~\ref{fig:transcor}. We also checked that our chain had converged, both by visual inspection and using the Gelmen-Rubin (G-R) statistic \citep{gelman,Ford2006}.}


\begin{table}
\begin{center}
\caption{Proposal parameters of the model used in our MCMC analysis}
\begin{tabular}{|l|l|}
\hline
$T_{c}$ & Time of mid transit\\

P & Period of planet\\

$\Delta F$ & Depth of transit\\

$T_{14}$ & Transit duration\\

b & Impact parameter\\

$K_{1}$ & Stellar radial reflex velocity\\

$\rm{T}_{\rm{eff}}$ & Effective temperature of the star\\

$[\frac{Fe}{H}]$ & Metallicity of the star\\

\hspace*{-7pt}$\left.\begin{tabular}{l|r|}
$\sqrt{e}\cos{\omega}$\\
$\sqrt{e}\sin{\omega}$
\end{tabular}\right\}$ & e=eccentricity, $\omega$ = argument of periastron \\

$\Delta F_{3.6}$ & Depth of secondary eclipse at 3.6$\mu$m\\

$\Delta F_{4.5}$ & Depth of secondary eclipse at 4.5$\mu$m\\
\hline
\end{tabular}
\label{params}
\end{center}
\end{table}

\begin{table}
\caption{Limb darkening coefficients
}\begin{center}
\begin{tabular}{|l|r|r|r|r|}
\hline
\multicolumn{1}{l}{Light Curve} &
\multicolumn{1}{c}{$a_{1}$} &
\multicolumn{1}{c}{$a_{2}$} &
\multicolumn{1}{c}{$a_{3}$} &
\multicolumn{1}{c}{$a_{4}$} \\
\hline
FTS              & 0.655 & $-$0.352 &  0.645 & $-$0.329\\
BUSCA (\emph{g} band)   & 0.433 &  0.208 &  0.496 & $-$0.300\\
BUSCA (\emph{r} band)   & 0.555 &  0.028 &  0.445 & $-$0.278\\
BUSCA (\emph{i} band)   & 0.641 & $-$0.267 &  0.640 & $-$0.338\\
\hline
\end{tabular}
\label{ld}
\end{center}
\end{table}

\subsection{Trend Functions and Aperture size}

\begin{figure}
\includegraphics[scale=0.5]{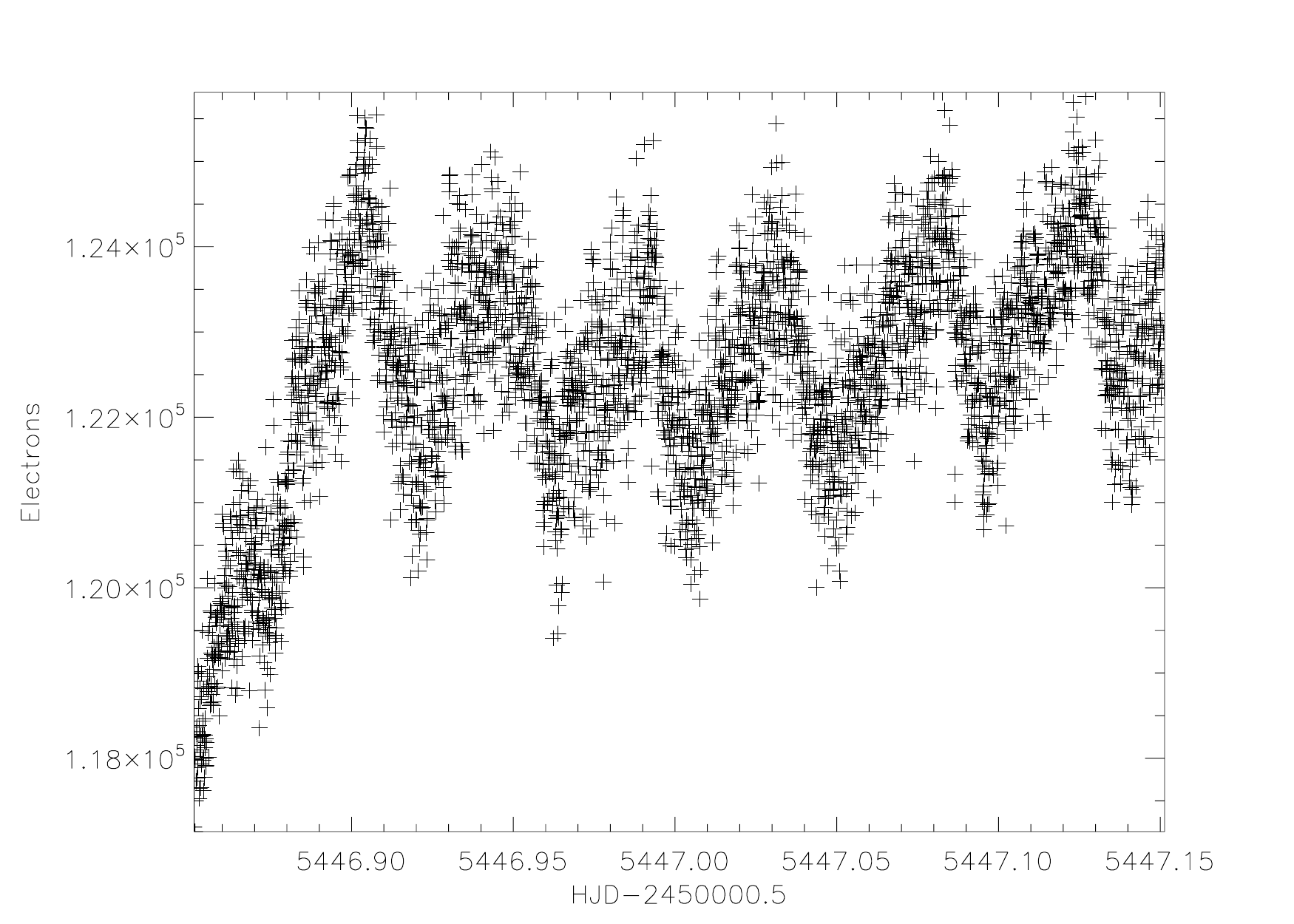}
\caption{The raw light curve of the $3.6\micron$ \emph{Spitzer} data extracted using and aperture of 2.4 pixels.}
\label{eglc1}
\end{figure}

\begin{figure}
\includegraphics[scale=0.5]{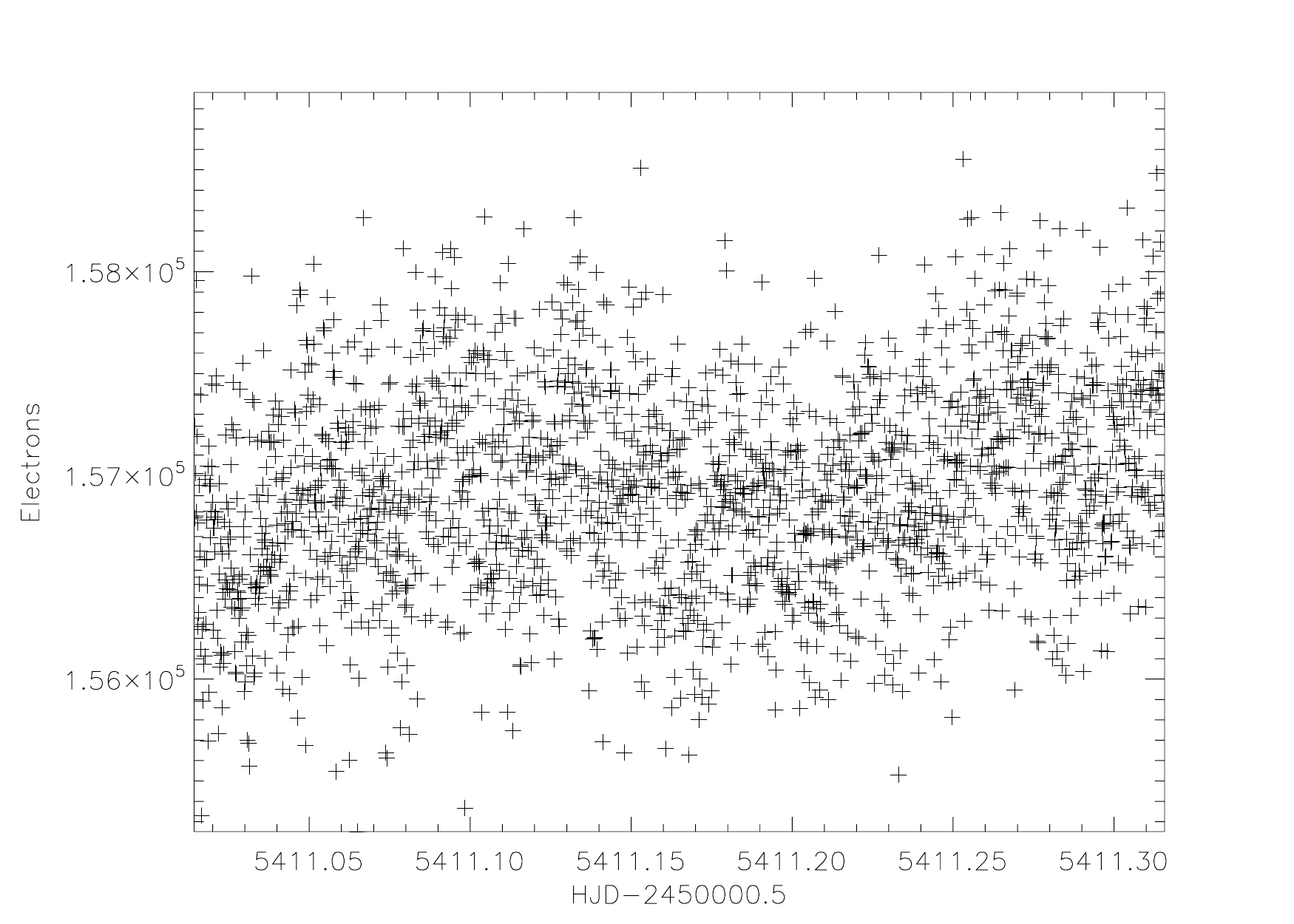}
\caption{The raw light curve of the $4.5\micron$ \emph{Spitzer} data extracted using an aperture of 2.4 pixels.}
\label{eglc2}
\end{figure}

\begin{figure*}
\includegraphics[scale=0.9]{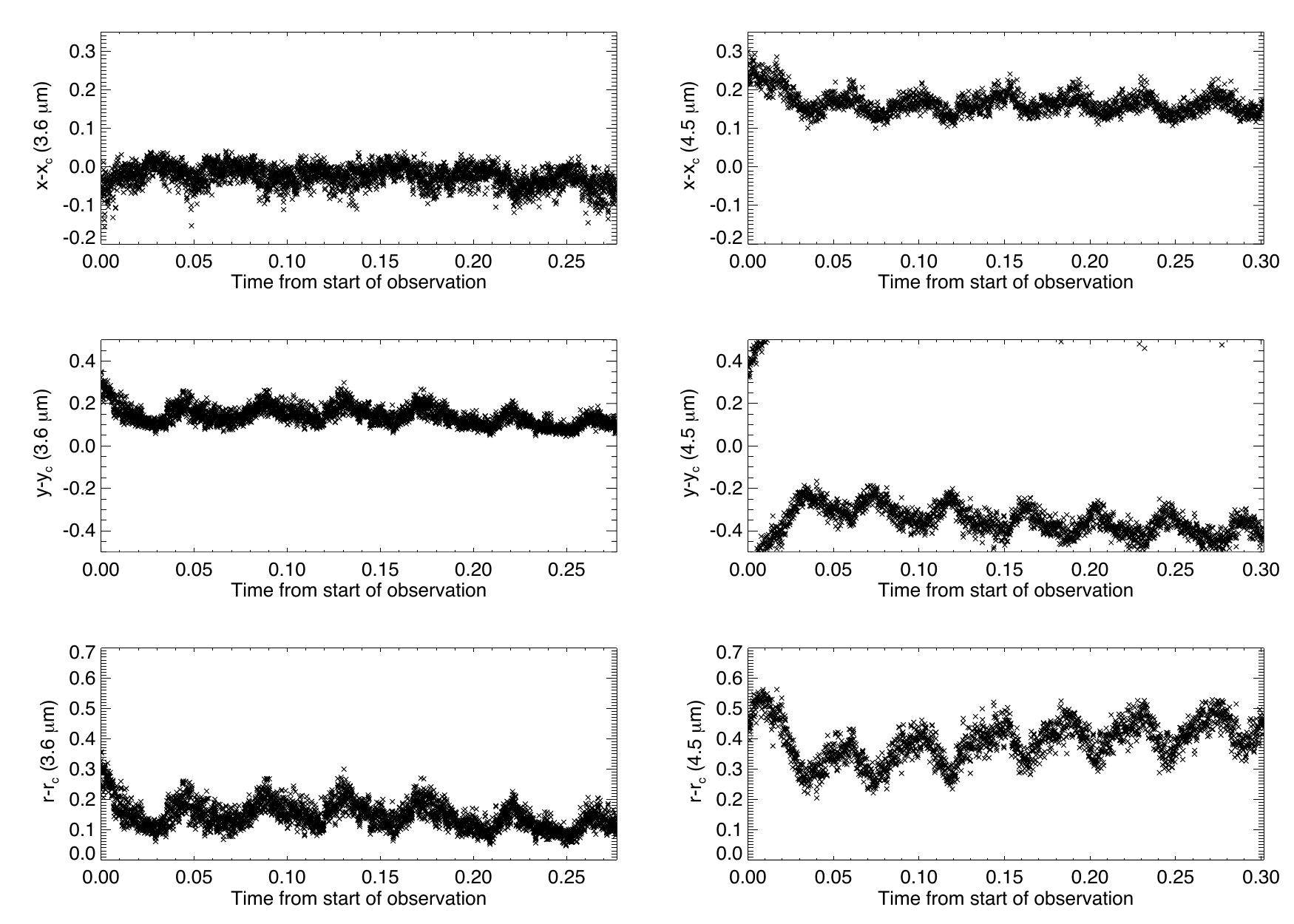}
\caption{The top, middle and bottom plots of the each column show the distance of the PSF from the nearest pixel centre in $x$, $y$ and radially in each of the measured wavelengths respectively.}
\label{three}
\end{figure*}

Figure \ref{eglc1} shows an example of the 3.6$\micron$ light curve produced
by the photometry in IRAF. There is a steep increase in the measured flux
during the first part of the observation. This occurs because the telescope
has slewed from its old position to its new position and is adjusting to a new
equilibrium. We exclude the data that precedes HJD=2455447.37, to remove the
major part of the initial ramp. It can be seen that there is a clear periodic
trend in the data. This is due to the variation in the position of the target
on the detector caused by flexure of the instrument as  an electric heater is
turned off and
on.\footnote{\emph{ssc.spitzer.caltech.edu/warmmission/news/21oct2010memo.pdf}}
The \emph{IRAC} detectors are known to exhibit inhomogeneous intrapixel
sensitivity  \citep[e.g.][]{knutson2008}, which means that different parts of
the detector are more or less sensitive than others. This, along with the PSF
movement, results in the measured flux varying depending on the position of
the PSF on the detector. Also, when small apertures are used pixelation occurs
due to the under-sampling of the PSF of the target  \citep{anderson2011b}.
These systematics will be accounted for in the trend functions as described
below.  Figure \ref{eglc2} shows an example of the 4.5$\micron$ data which is
less affected by these systematics even though (as it can be seen from Figure
\ref{three}) the radial motion of the PSF is greater at 4.5$\micron$ than at
3.6$\micron$.

The general form of the trend functions that were used in our analysis is
\begin{equation}
\label{trend}
\begin{array}{ll}
\Delta f = &  a_0 + a_x\Delta x + a_y\Delta y + a_{xy}\Delta x\Delta y +
a_{xx}\Delta x^2 \\ 
& + a_{yy}\Delta y^2 + a_t\Delta t,\\
\end{array}
\end{equation}
\noindent where $\Delta f = f - \hat{f}$ is the stellar flux relative to its weighted mean, $\Delta x = x - \hat{x}$ and $\Delta y = y - \hat{y}$ are the coordinates of the point spread function of the target centre relative to their weighted means, $\Delta t$ is the time since the beginning of the observation, and $a_0$, $a_x$, $a_y$, $a_{xx}$, $a_{yy}$ and $a_t$ are coefficients which are free parameters in the MCMC analysis \citep{anderson2011b}. \reff{For each set of trial lightcurve model parameters we calculate the residuals from the model and then calculate the coefficients of the detrending model using singular value decomposition applied to the entire data set.} Initially, a linear-in-time and quadratic-in-space trend function was used on all 21 apertures to fit the secondary eclipse data. The RMS of the residuals was used to determine the optimal aperture size. Once this was determined, combinations of no, linear and quadratic trend functions in time and space were used on the best aperture to determine the best fitting trend function.

Initially this decorrelation was conducted using the positions measured by the
1-dimensional Gaussian fit to the target. We also attempted to remove the
trends in the data by decorrelating against the radial position (radial
distance from the centre of the nearest pixel) instead of the $x$ and $y$
positions independently. The general trend function for the radial
decorrelation is,
\begin{equation}
\label{rtrend}
\Delta f = b_0 + b_1r + b_2r^{2} + b_t\Delta t,
\end{equation}
\noindent where $b_0,b_1,b_2,b_t$ are free parameters in the MCMC analysis and
$r$ is the radial distance from the centre of the nearest pixel centre.
 A third method that was attempted was to use target positions in the trend
functions measured  by fitting a two dimensional circular Gaussian of fixed
full width half maximum (1.39 pixels in channel 1 and 1.41 pixels in channel
2) to a small region of the images containing the target.

To determine which trend function gave better results we used the Bayesian Information Criterion (BIC) \citep{schwarz1978},
\begin{equation}
\rm{BIC} = \chi^{2} +\emph{k}\ln(\emph{n})
\label{bic}
\end{equation}
where \emph{k} is the number of free parameters and \emph{n} is the number of data points. This method of determining how complicated a model to use only accepts a higher order trend function if the fit improves $\chi^{2}$ by $\ln(\emph{n})$ or better for each additional free parameter. 

\begin{figure*}
\includegraphics[scale=0.8]{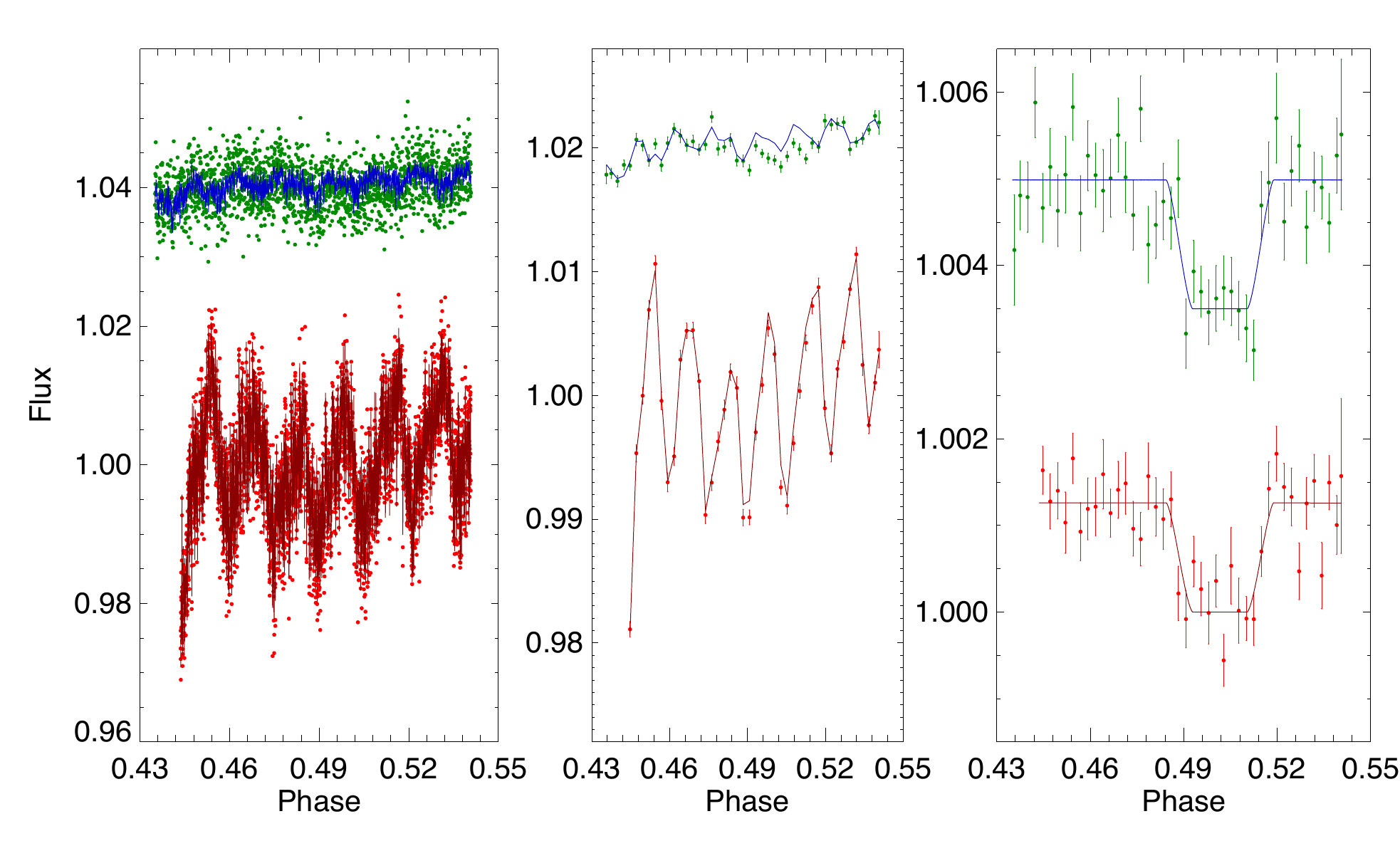}
\caption{(Left): The raw light curves with the trend functions, the upper points are the channel 2 data and the lower points are the channel 1 data, the  solid lines are the trend functions for each data set. (Middle) Binned light curves with trend models. (Right) The binned light curve with trend function removed and best fitting eclipse models (solid lines). The secondary eclipse can clearly be seen in both channels. }
\label{res}
\end{figure*}

Using the RMS of the residuals it was found that the best aperture to use was 2.4 pixels in both channels.  The system parameters are negligibly affected by the choice of aperture radius around this value. It was also found that the RMS of the residuals to the channel 1 data was  marginally lower when using the position measurements measured by the 2D circular Gaussian method as opposed to 1D Gaussian position measurements (0.002995 compared to 0.003054). The channel 2 data gave consistent RMS no matter the position measurement used. The system parameters were consistent no matter which position measurement system were used. The results shown in Figure \ref{res} and Table \ref{tab:mcmc} are those using the 2D circular Gaussian method, extracted from the 2.4 pixel aperture and trend functions as described below. We found that the radial decorrelation gave a worse fit to our data compared to that of $x$ and $y$ decorrelation ($\chi^2$ worse by $\sim$ 3000 at 3.6$\micron$ and $\sim$ 400 at 4.5$\micron$). 

Using equation (\ref{bic}) it was found that the quadratic-in-space with no time trend function gave the best fit to the data in channel 1 and that the linear-in-space with no time trend function gave the best fit to the data in channel 2. \reff{It was found that the addition of the quadratic term for the spacial decorrelation improved our BIC by $\sim200$ in channel 1 and less than $\sim10$ for more complicated models in both channels.} \reff{We also detrended our data based only on the out-of-eclipse points to see if this affected our measured eclipse depths. It was found that the eclipse depths were consistent with our previous decorrelation.} 


\begin{table*} 
\caption{System parameters from our MCMC analysis} 
\label{tab:mcmc}
\begin{tabular}{lllclr}
\hline
Parameter & Symbol (unit) & Value
\\ 
\hline 
\\
Orbital period & $P$ (d) &  2.756611 ${\pm 0.000008}$\\[0.5ex]
 
Epoch of mid-transit (BJD, TDB) & $T_{\rm c}$ & 2455424.10899 ${\pm 0.00012}$\\[0.5ex]
 
Transit duration (from first to fourth contact) & $T_{\rm 14}$ (d) & 0.097 ${\pm 0.002}$\\[0.5ex]
 
Duration of transit ingress $\approx$ duration of transit egress & $T_{\rm 12} \approx T_{\rm 34}$ (d) & 0.024 ${\pm0.002}$\\[0.5ex]
 
Planet-to-star area ratio & $\Delta F=R_{\rm P}^{2}$/R$_{*}^{2}$ & 0.0103 ${\pm 0.0003}$\\[0.5ex]
 
Impact parameter & $b$ & 0.82 ${\pm 0.02}$\\[0.5ex]
 
Orbital inclination & $i$ ($^\circ$) \medskip & 82.9 ${\pm 0.4}$\\[0.5ex]
 
Semi-amplitude of the stellar reflex velocity & $K_{\rm 1}$ (km s$^{-1}$) & 0.138${\pm 0.002}$\\[0.5ex]

Centre-of-mass velocity & $\gamma$ (km s$^{-1}$) \medskip & 8.4593 ${\pm 0.0001}$\\[0.5ex]

Argument of periastron & $\omega$ ($^\circ$) & $-$90$ ^{+ 200}_{- 20}$\\[0.5ex]

& $e\cos\omega$ & $-$0.0004 ${\pm 0.0007}$\\[0.5ex]

& $e\sin\omega$ & $-$0.0011 $^{+ 0.0023}_{- 0.0110}$\\[0.5ex]

Orbital eccentricity & $e$ & 0.00283 $^{+ 0.00965}_{- 0.00221}$\\[0.5ex]
 
Phase of mid-occultation & $\phi_{\rm mid-occultation}$ & 0.4998 ${\pm 0.0005}$\\[0.5ex]
 
Occultation duration & $T_{\rm 58}$ (d) & 0.097 ${\pm 0.002}$\\[0.5ex]
 
Duration of occultation ingress $\approx$ duration of occultation egress & $T_{\rm 56} \approx T_{\rm 78}$ (d) \medskip & 0.024${\pm 0.002}$\\[0.5ex]
 
Star mass & $M_{\rm *}$ ($M_{\rm \odot}$) & 1.10${\pm 0.03}$\\[0.5ex]
 
Star radius & $R_{\rm *}$ ($R_{\rm \odot}$) & 1.29${\pm0.05}$\\[0.5ex]
 
Star surface gravity & $\log g_{*}$ (cgs) & 4.26 ${\pm 0.03}$\\[0.5ex]
  
Star density &  $\rho_{\rm *}$ ($\rho_{\rm \odot}$) & 0.52${\pm 0.06}$\\[0.5ex]
 
Star effective temperature & $T_{\rm eff}$ (K) & 6000${\pm 100}$\\[0.5ex]
 
Star metallicity & {[Fe/H]} \medskip &$-$0.02${\pm0.09}$\\[0.5ex]
 
Planet mass & $M_{\rm P}$ ($M_{\rm Jup}$) & 1.03${\pm0.02}$\\[0.5ex]
 
Planet radius & $R_{\rm P}$ ($R_{\rm Jup}$) & 1.27 ${\pm 0.07}$\\[0.5ex]
 
Planet surface gravity & $\log g_{\rm P}$ (cgs) & 3.16 ${\pm 0.04}$\\[0.5ex]
 
Planet density & $\rho_{\rm P}$ ($\rho_{\rm J}$) & 0.50 ${\pm0.08}$\\[0.5ex]
 
Semi-major axis & $a$ (AU)  & 0.0398 ${\pm 0.0003}$\\[0.5ex]

Occultation depth at $3.6\mu$m & $\Delta F_{3.6}$ & $0.00126 \pm 0.00013$ \\[0.5ex]

Occultation depth at $4.5\mu$m & $\Delta F_{4.5}$ & $0.00149 \pm 0.00016$ \\[0.5ex]
 
Planet equilibrium temperature (full redistribution)$^*$  &$T_{\rm P, A=0,f=1}$ (K) & 1623 ${\pm 43}$\\[0.5ex]

Planet equilibrium temperature (day-side redistribution)$^*$  &$T_{\rm P, A=0,f=2}$ (K) &  ${1930 \pm 51}$\\[0.5ex]

Planet equilibrium temperature (instant reradiation)$^*$ &$T_{\rm P, A=0,f=\frac{8}{3}}$ (K) &  ${2074\pm 55}$\\[0.5ex]
\hline 
\end{tabular} 
\\$^*$ where A is the albedo, f=1 is defined as full redistribution, \\f=2 is day-side redistribution, and f=$\frac{8}{3}$ is instant reradiation as in \citet{smith2011}
\end{table*}

\section{Results and Discussion}

\subsection{Eclipse Depths and Brightness Temperatures}
We find that the eclipse depths at $3.6\micron$ and $4.5\micron$ are $0.00126
\pm 0.00013$ and $0.00149 \pm 0.00016$, respectively. These eclipse depths
correspond to brightness temperatures of $1825\pm80$K and $1725\pm89$K. To
find these blackbody temperatures the expected flux ratios were calculated
using Planck functions at different temperatures for the planet and synthetic spectra
from stellar models \citep{stars} for the star. These flux ratios were then
integrated over the \emph{Spitzer} band passes to calculate the expected
measured flux ratio. The temperatures above correspond to the best fitting
Planck function temperature to the individual eclipse depths. The errors were
calculated using a simple Monte Carlo method. These temperatures suggest that,
on average, the emission at mid-infrared wavelengths from the irradiated
hemisphere of WASP-26b is consistent with the spectrum of an isothermal
atmosphere, with the possibility of a weak thermal inversion within the
uncertainties on the brightness temperatures.

\subsection{Atmospheric Analysis}

We model the day-side emergent spectrum of the hot Jupiter WASP-26b using the
atmospheric modeling and retrieval technique of
\citet{madhu2009,Madhusudhan2010}. The model computes line-by-line radiative
transfer in a plane-parallel atmosphere in local thermodynamic equilibrium,
and assumes hydrostatic equilibrium and global energy balance. The
pressure-temperature ($P$-$T$) profile of the atmosphere and 
the chemical
composition, i.e. the sources of molecular line opacity, are input parameters
to the model. The model atmosphere includes the major sources of opacity
expected in hot, hydrogen-dominated atmospheres, namely, molecular absorption
due to H$_{2}$O, CO, CH$_{4}$, and CO$_{2}$, and continuum opacity due to
H$_{2}$-H$_{2}$ collision-induced absorption (CIA). Our molecular line-lists
are discussed in \citet{madhu2009} and \citet{smith2012}. Given a photometric
or spectral dataset of thermal emission from the planet, we explore the space
of atmospheric chemical composition and temperature structure to determine the
regions in model space that explain, or are excluded by, the data
\citep[e.g.][]{Madhusudhan2011}. In the present case, however, the number of
available data points ($N = 2$) are far below the number of model parameters
($N = 10$), implying that a unique model fit to the data is not feasible.
Consequently, we nominally fixed the chemical composition of the models to
that obtained with solar elemental abundances in thermochemical equilibrium
\citep[e.g.][]{burrows1999,madhu12} for a given thermal profile, and explored
the space of thermal profiles, with and without thermal inversions, that might
explain the data.  

\begin{figure}
\begin{center}
\includegraphics[scale=0.5]{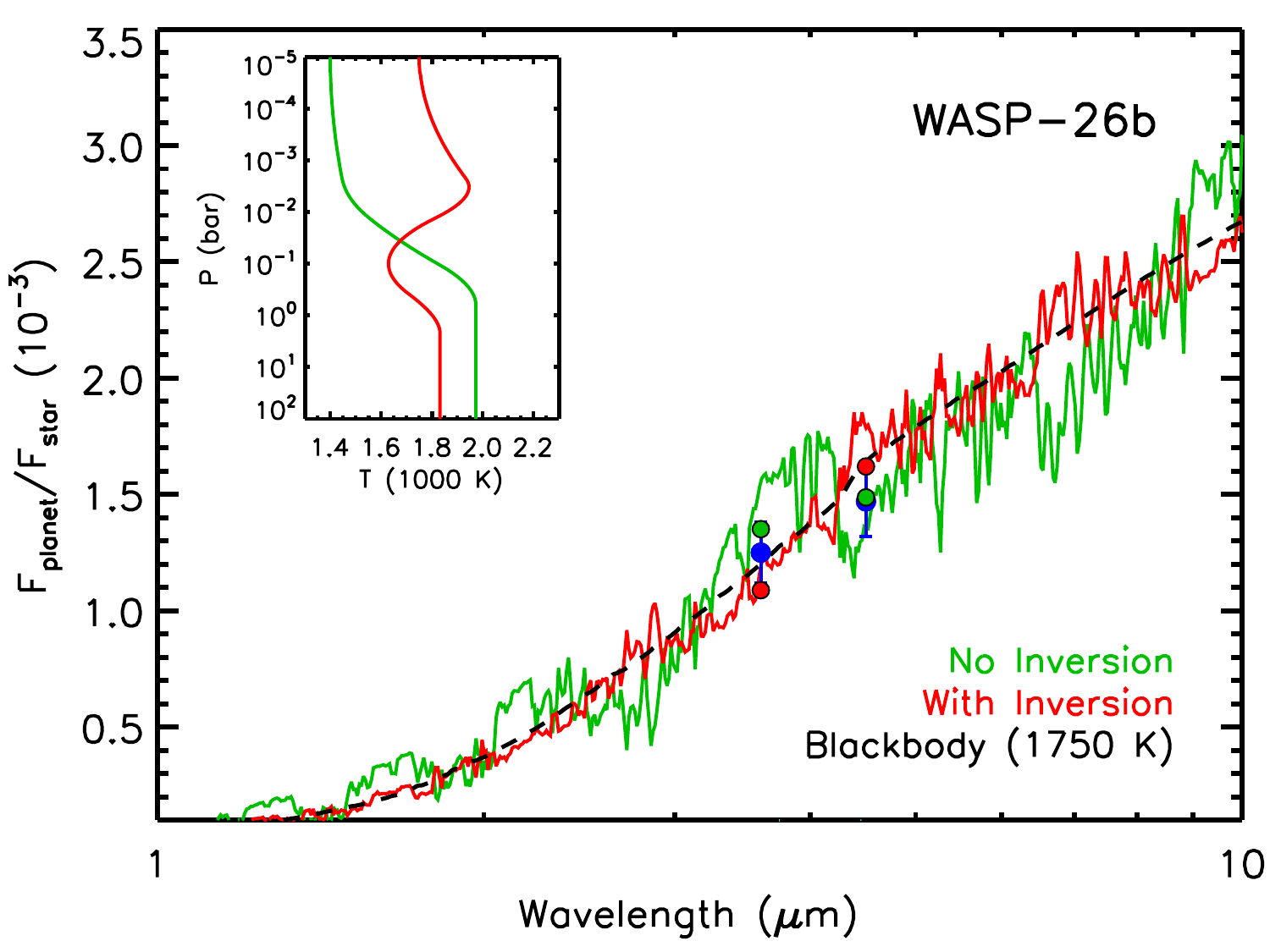}
\caption{Spectral energy distribution of WASP-26b relative to that of its host star. The blue circles with error bars are our best-fitting occultation depths. The green line is a model-atmosphere spectrum, based on a model which assumes solar abundances in thermochemical equilibrium and lacks a temperature inversion, and the dark red line is a model with a temperature inversion. The band-integrated model fluxes are indicated with circles of the corresponding colours. The dashed black line shows a planetary black body model with a temperature of 1750K. Inset: temperature-pressure profiles for our models.}
\label{model}
\end{center}
\end{figure}

Figure \ref{model} shows the $3.6\micron$ and $4.5\micron$ data along with
model spectra of atmospheres with and without a thermal inversion, and a
blackbody model. All three models shown allow for very efficient day-night
redistribution. We find that both our planet-star flux ratios can be explained
by a planetary blackbody at around 1750 K. Consequently, the data are
consistent with an isothermal atmosphere. However, an isothermal temperature
profile may be unphysical in radiatively efficient atmospheres at low optical
depth \citep[e.g.][]{hansen2008}. A temperature profile with a non-zero
thermal gradient, with or without a thermal inversion, may be more plausible.
As shown in Fig.~\ref{model}, the two data are fit almost equally well by
models with and without a thermal inversion, as shown by the red and green
models, respectively. Further occultation depths measured at different
wavelengths are required to break the degeneracies between the models and to determine the true nature of the atmosphere. It can be seen in Fig.~\ref{model} that there are some differences between the models with and without a thermal inversion at $1.25\micron$ (J band), $1.65\micron$ (H band) and $2.2\micron$ (K band). These wavelengths are accessible from the ground, so with measurements of the occultation depth at these wavelengths it may be possible to break the degeneracies between these models. Hubble Space Telescope WFC3 observations covering the wavelength range $1-1.7\micron$ can also be used to detect spectral features due to water either in emission or absorption, and so distinguish between models with and without a thermal inversion \citep{madhu12, swain12}. We emphasize that we have only presented two possible models here that represent the average properties of the irradiated hemisphere of WASP-26b. With additional data other parameters of the models such as composition can be explored.

\subsection{Activity-Inversion Relation}

\citet{Knutson2010} (hereafter K10) presented results which suggest  that
planets without thermal inversions orbit active stars, and those with
inversions orbit inactive stars. This may be due to photodissociation of the
opacity source in the upper atmosphere of the planet by the  UV flux from the
active stars (K10). It is known that solar-like stars have activity cycles on
time scales of approximately 10 years. The \citet{Duncan1991} catalogue of
$S_{HK}$ activity measurements taken at the Mount Wilson Observatory was used
to examine to what extent the activity of a star changes on short time scales
(order of months) and long time scales (order of years). The aim was to
determine if the variability in activity  of the stars in the K10 sample was
such that, in the time between the occultation observation and the measurement
of $\log{R'_{HK}}$, the activity of the star can change enough to affect the
interpretation of this activity-inversion relation. Recently \citet{wasp3}
showed that the activity of WASP-3 changed from $\log{R'_{HK}}= -4.95$ (less
active) to $\log{R'_{HK}}=-4.8$ (more active) between 2007 and 2010. It has
been shown by \citet{atmosphere} that the time scale for models of hot Jupiter
atmospheres to go from their initial conditions to a statistical steady state
was $\sim$ 20 days. This suggests the time scale of hot Jupiter atmosphere
variability is much shorter than the time scale of stellar activity
variability. More detailed modelling and additional observations are  required
to better understand whether variations in the UV irradiation can produce
observable changes in the eclipse depths for planets near the boundary between
atmospheres with and without strong thermal inversions.

\begin{figure*}
\includegraphics[scale=0.3]{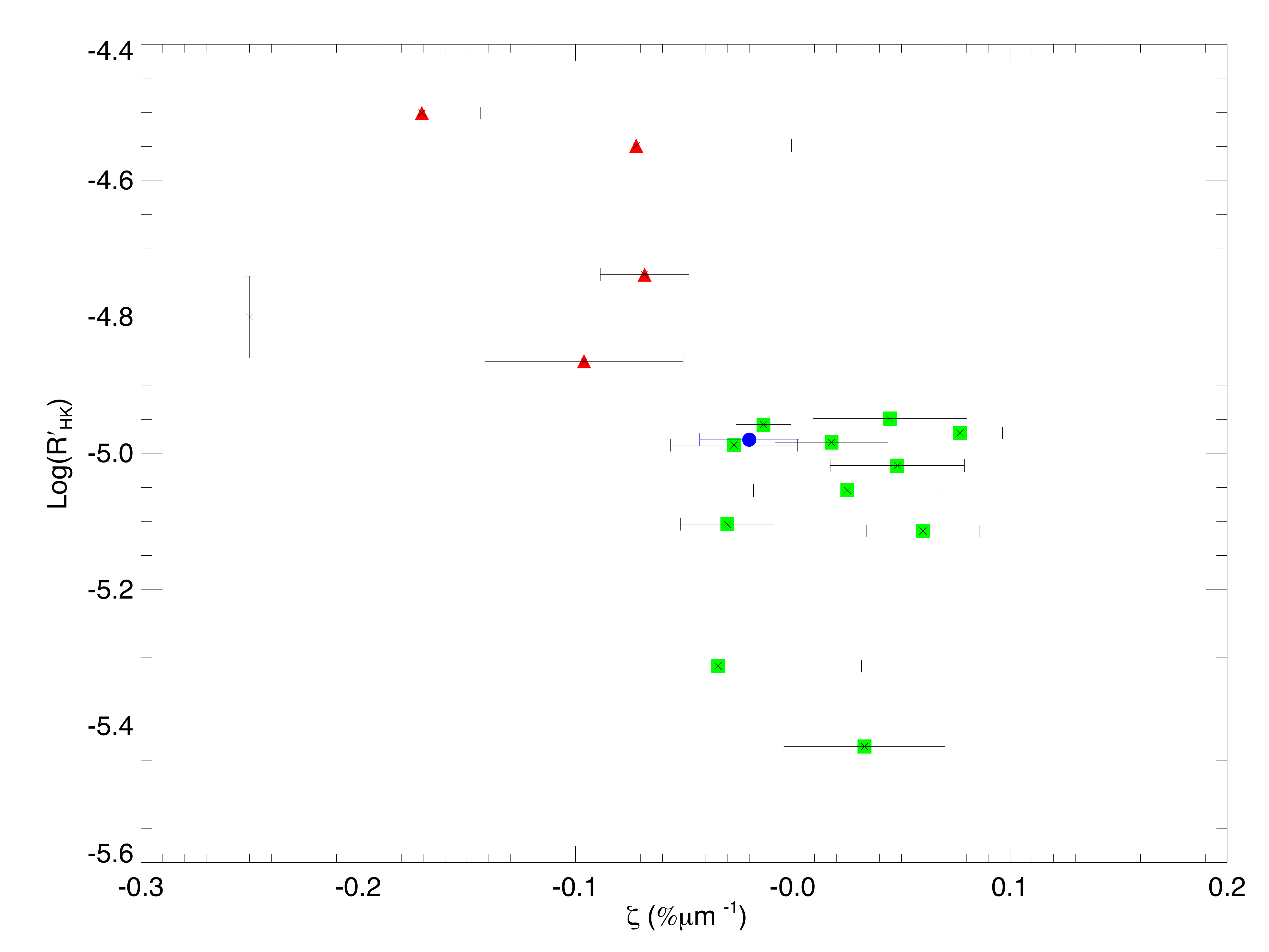}
\caption{Activity-inversion plot for the stars in \citet{Knutson2010}. Points on the left of the dotted line (triangles) are non-inverted planets around active stars and those on the right of the dotted line (squares) are inverted planets around inactive stars. The point on the left hand side of the plot shows the typical change in $\log{R'_{HK}}$ (season-to-season), assuming measurements over several nights. The blue circle is WASP-26. The stars are listed in table \ref{kstars}.}
\label{knutson_plot}
\end{figure*}

 We converted the $S_{HK}$ measurements in \citet{Duncan1991} to
$\log{R'_{HK}}$ using the method described by \citet{Noyes1984}. A look-up
table based on $\log{R'_{HK}}$ and B--V  colour for the stars in the
\citet{Duncan1991} catalogue was then constructed. Using this table, the
within-season variation of $\log{R'_{HK}}$ of the stars was used as a measure
of the short term variability in $\log{R'_{HK}}$ and the season-to-season
variation in $\log{R'_{HK}}$ as a measure of the long term variation in
$\log{R'_{HK}}$. This look up table was then used to estimate the variation
in $\log{R'_{HK}}$ for the stars of K10 based on their B--V colour.
 It was found that the short term variability was always
$\le0.02$ dex and the long term variability was between 0.02 and 0.06 dex.
This suggests that the variation in $\log{R'_{HK}}$ is not large enough on
either short nor long term time scales to change the interpretation of K10.
However, this may blur the boundary between the two classes of planets. The
error bar shown in Figure \ref{knutson_plot} is the typical change in
activity, assuming the spectra are measured over several nights. It is
possible for stars to vary by much more than this amount over their rotation
period \citep[e.g.][]{Dumu2012}. This short time scale variation will move the
star on the diagram but this may not reflect  changes in UV irradiation. The
value of $\log{R'_{HK}} = -4.98$ for WASP-26 used in this analysis is taken from
\citet{anderson2011}.

We compiled updated values of $R_{p}/R_{\star}$ and the  secondary eclipse
depths for the stars in the K10 sample. Figure 5 of K10 was then replotted,
this is shown in Figure \ref{knutson_plot}. We include on this plot WASP-26b.
As can be seen from Figure \ref{knutson_plot}, it seems to lie very close to
the boundary between the two classes. Using the convention as in
\citet{anderson2011a} the abscissa value for WASP-26b is $\zeta =-0.020 \pm
-0.023\% \micron^{-1}$, where $\zeta$ is the gradient of the measurements at
$3.6\micron$ and $4.5\micron$, i.e. $ \Delta F_{3.6} - \Delta
F_{4.5}/({-0.9\micron})$, minus the gradient of the blackbody that is the
best--fit to the two measurements. The theory behind this is that at
$4.5\micron$ there are opacity sources that are not present at $3.6\micron$
($\rm{CO}$ and $\rm{H}_2\rm{O}$) \citep{Madhusudhan2010}. The $4.5\micron$
data probes a higher region of the atmosphere compared to the $3.6\micron$
data. This suggest that if the brightness temperature at  $4.5\micron$ is
greater than that at  $3.6\micron$ then there is likely to be a thermal 
 inversion in the atmosphere.

\begin{table*}
\caption{Stars in Figure \ref{knutson_plot}}
\label{kstars}
\begin{tabular}{l|r|r|l|l|l}
\hline
\multicolumn{1}{c}{Star} & 
\multicolumn{1}{c}{Log(R$'_{HK}$)} & 
\multicolumn{1}{c}{$\zeta$ value}&
\multicolumn{1}{c}{$R_{P}/R_{\star}$}&
\multicolumn{1}{c}{Eclipse depth in channel 1}&
\multicolumn{1}{c}{Eclipse depth in channel 2}\\
\hline
HD189733 &  $-$4.501$^{*}$ & $-$0.1707  $\pm$ 0.0271 & \citet{carter2010} & \citet{charbon2008} & \citet{charbon2008}\\ 
TRES-3  &      $-$4.549$^{*}$ & $-$0.0721 $\pm$ 0.0715 & \citet{southworth2011} & \citet{fressin2010} & \citet{fressin2010} \\
TRES-1    &    $-$4.738$^{*}$ & $-$0.0682 $\pm$ 0.0204 & \citet{Southworth2008} & \citet{Knutson2010} & \citet{char2005}\\
WASP-4  &      $-$4.865$^{*}$ & $-$0.0961 $\pm$ 0.0457 & \citet{southworth2009a} & \citet{beerer2011} &  \citet{beerer2011}\\
XO-2    &          $-$4.988$^{*}$ & $-$0.0271 $\pm$ 0.0292 & \citet{southworth2010} & \citet{mach2009} & \citet{mach2009}\\
TRES-2  &       $-$4.949$^{*}$ &    0.0447   $\pm$ 0.0354 & \citet{southworth2011} & \citet{odon2010} & \citet{odon2010}\\
XO-1    &          $-$4.958$^{*}$ & $-$0.0135 $\pm$ 0.0127 & \cite{burke2010} & \citet{mach2008} &  \citet{mach2008}\\
HAT-P-1 &       $-$4.984$^{*}$ &    0.0178   $\pm$ 0.0260 & \citet{Southworth2008} &\citet{torodov2010} &\citet{torodov2010}\\
HD209458 &   $-$4.970$^{*}$ &    0.0770   $\pm$ 0.0194 & \citet{Southworth2008} & \citet{knutson2008} & \citet{knutson2008}\\
TRES-4  &       $-$5.104$^{*}$ & $-$0.0301 $\pm$ 0.0216 & \citet{southworth2012a} & \citet{Knutson2009a} & \citet{Knutson2009a}\\
COROT-1 &     $-$5.312$^{*}$ & $-$0.0343 $\pm$ 0.0660 & \citet{southworth2011} & \citet{dem2011}& \citet{dem2011}\\
WASP-1    &     $-$5.114$^{*}$ &    0.0599   $\pm$ 0.0259& \citet{Southworth2008} & \citet{wheat2010} & \citet{wheat2010} \\
WASP-2  &      $-$5.054$^{*}$ &    0.0251   $\pm$ 0.0432 & \cite{southworth2010a}& \citet{wheat2010} & \citet{wheat2010}\\
WASP-18 &     $-$5.430$^{*}$ &    0.0332   $\pm$ 0.0176& \citet{southworth2009b} & \citet{Maxted2012} & \citet{Maxted2012}\\
HAT-P-7    &    $-$5.018$^{*}$ &    0.0481   $\pm$ 0.0309& \citet{southworth2011}& \citet{chris2010} & \citet{chris2010}\\
WASP-26 &     $-$4.98$^{**}$ &  $-$0.0200 $\pm$ 0.0229 & This paper & This paper & This paper\\
\hline
\end{tabular}
$^{*}$ Log(R$'_{HK}$) value from \citet{Knutson2010}\\
$^{**}$ Log(R$'_{HK}$) value from \citet{anderson2011}\\
\end{table*}

\subsection{Ecentrictiy}
From secondary eclipse measurements it is also possible to constrain the eccentricity of the orbit from timing of the secondary eclipse relative to transit. We find that the eccentricity of the orbit is small (e $= 0.0028 ^{+ 0.0097}_{- 0.0022}$), which is consistent with a circular orbit at the $1\sigma$ level. We find a $3\sigma$ upper limit on the eccentricity of the planet's orbit of $0.0399$ which is similar to \citet{anderson2011} $3\sigma$ upper limit of 0.048.

\section{conclusion}
In this paper we present new warm \emph{Spitzer} photometry of WASP-26 at
$3.6\micron$ and $4.5\micron$ along with new transit photometry taken in the
\emph{g,r} and \emph{i} bands. We report the first detection of the
occultation of WASP-26b with eclipse depths at $3.6\micron$ and $4.5\micron$
of $0.00126 \pm 0.00013$ and $0.00149 \pm 0.00016$ respectively which
correspond to brightness temperatures of $1825\pm80$K and $1725\pm89$K. Our
analysis shows that the atmosphere of WASP-26b is consistent with an
isothermal atmosphere with the possibility of a weak thermal inversion (within
the uncertainties on the brightness temperatures). If the K10
activity-inversion relation holds for WASP-26b, then we would expect it to
host a thermal inversion.  More secondary eclipse data at different
wavelengths, particularly near-IR secondary eclipse depths near the peak of
the planet's SED, will be able to better constrain the true nature of the atmosphere
of WASP-26b.

\section{acknowledgements}
DPM and JTR acknowledge the financial support from STFC in the form of Ph.D. studentships. JS acknowledges financial support from STFC in the form of an Advanced Fellowship. This work is based in part on observations made with the Spitzer Space Telescope, which is operated by the Jet Propulsion Laboratory, California Institute of Technology, under a contract with NASA.  Support for this work was provided in part by NASA through awards issued by JPL/Caltech and by the Planetary Atmospheres Program, grant NNX12AI69G. NM acknowledges support from the Yale Center for Astronomy and
Astrophysics (YCAA) at Yale University through the YCAA prize
postdoctoral fellowship.

\bibliographystyle{mn2e}
\bibliography{w26}
\end{document}